\documentclass[prb,twocolumn,aps,showpacs,preprintnumbers]{revtex4}
\usepackage{graphicx}
\usepackage{subfigure}

\usepackage[subfigure,caption2]{ccaption}

\usepackage{dcolumn}
\usepackage{bm}
\usepackage{txfonts}
\usepackage{amssymb}

\begin{document}

\newcommand{\om}{\omega}

\title{Current response of ac\textendash driven nanoelectromechanical systems in single-electron tunneling regime}

\author{G.~Labadze and Ya.~M.~Blanter}
 \affiliation{Kavli Institute of Nanoscience, Delft University of
 Technology, Lorentzweg 1, 2628 CJ Delft, The Netherlands}

\date{\today}

\begin{abstract}
We investigate electric current in a single-electron tunnelling device weakly coupled to an ac-driven underdamped harmonic nanomechanical oscillator. In the linear regime, the current can respond to the external frequency in a resonant as well as in an anti-resonant fashion. The main resonance is accompanied by an additional resonance at a half of the external frequency.
\end{abstract}

\maketitle

\section{Introduction}

During last decade the nanoelectromechanical systems (NEMS) were widely studied because of richness of physical effects and their broad spectrum of the practical functionalities\cite{Clelandbook}. With small inertial masses and high frequencies of the nanomechanical resonators, together with the ultra-sensitive mechanical displacement detection capabilities of electronic devices used as detectors, NEMS have great promise for application in the force microscopy and mass-sensing \cite{Tang+Rouks07,Metzger08,Yang06,Yang04,Munday09}. Many of the applications currently envisioned, such as switches, relays, and actuators, are directly related to the effect of mechanical motion on the electric properties of the system.

Recently, the first observation of quantum superposition of states of a mechanical resonator has been reported \cite{Martinis}. Bottom-up fabricated NEMS devices are prospective candidates for investigation of quantum effects in NEMS, since they have frequencies in GHz range, and these frequencies are tunable by the gate voltage\cite{Sapmaz,Sazonova}. These devices, made of suspended carbon nanotubes, function in single-electron tunneling (SET) regime.
The coupling between the charge of the SET and mechanical degrees of freedom is provided by the position dependence of the capacitance between
suspended beam and the underlying gate, and of the tunnel rates between the reservoirs and the SET island. It was theoretically predicted previously \cite{Usmani+Blanter07} that for certain energy-dependent tunnel rates even at weak coupling between the SET and mechanical motion the latter can considerably
influence electric current --- the phenomenon of strong mechanical feedback. At strong coupling (see Ref.\onlinecite{Bruder06}), the effect of the mechanical motion is even stronger, and the physics is dominated by polaronic effect. This leads to multistabilities and switching in the regime of $\om_0 \ll \Gamma$, where $\om_0$ and $\Gamma$ are the resonator frequency and the typical tunnel rate, respectively\cite{Martin}. For $\om_0 \gg \Gamma$, one observes Franck-Condon effect\cite{Flensberg03+Mirta05,Jarillo}.

Recently, careful experiments on electric transport in suspended carbon nanotubes have been performed\cite{Zant09}. In addition to source, drain and gate electrodes, an antenna emitting ac radiation was introduced close to the vibrating beam. A variety of interesting results have been obtained, including resonances, switching events, and gate voltage dependence of the quality factor of the mechanical resonator. There are indications that strong coupling regime may have been observed, at least in some parameter range.

In this Article, we consider simultaneous effects of ac and stochastic forces on transport in SET-based NEMS. A similar problem was previously considered in Ref. \onlinecite{Bruder04}. Our manuscript is different from Ref. \onlinecite{Bruder04} in several respects: We use a different theoretical formulation. In contrast to Ref. \onlinecite{Bruder04}, we consider the energy dependent tunneling rates (in addition to the energy dependence given by Fermi functions).We also calculate the dependence of the electric current on the driving frequency. We find that the regime of strong mechanical feedback, which was discussed in reference \onlinecite{Usmani+Blanter07}, disappears already at weak ac force. The main effect of the ac force is to create resonances when the external frequency coincides with $\om_0$ and $\om_0/2$. These resonances can enhance as well as suppress (anti-resonance) the current. We find that in some cases a resonance at $\om_0$ can be accompanied by an anti-resonance at $\om_0/2$ and vice versa.

\section{model and theory}

\begin{figure*}[t]
\centering
     \subfigure[]{\label{setup}\includegraphics[width=.4\textwidth,trim = 0mm 15mm 105mm 0mm, clip]{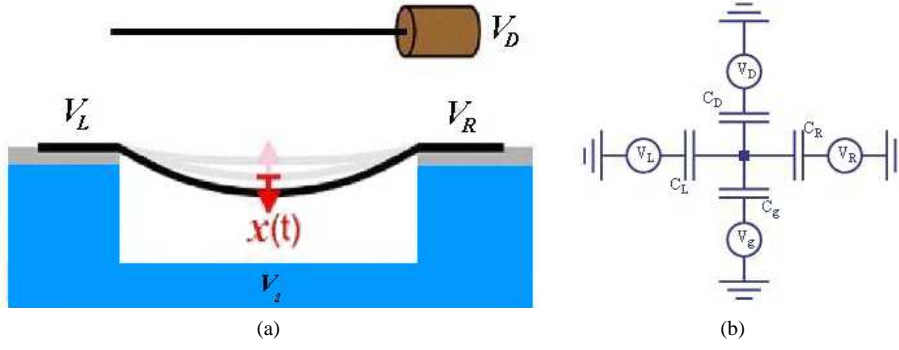}}
     \subfigure[]{\label{scheme}\includegraphics[width=.28\textwidth,trim = 150mm 15mm 0mm 0mm, clip]{setup.eps}}
\caption{
(a) Picture of the setup: a beam is suspended over two electrodes can oscillate freely. The beam is capacitively coupled to the gate electrode. Bias voltage $V_b=V_L-V_R$ is applied to the electrodes, and the beam is driven by the ac signal applied to the antenna. (b) Schematic version of setup: the dot in the center corresponds to the beam which is connected to four voltage sources via corresponding capacitances.
}
\label{setup-scheme}
\end{figure*}

The system we study is an SET device coupled to a harmonic oscillator with the frequency $\om_0$, mass $M$ and the quality factor $Q\gg 1$. The most obvious specific realization is a suspended beam (carbon nanotube) \cite{Sapmaz}, see Fig. \ref{setup-scheme}. The beam is suspended between two leads; the bias voltage $V_b=V_L-V_R$ is applied. The beam thus at the same time is an island where the electron charge is quantized.

The island is capacitively coupled to an underlying gate at which voltage $V_g$ is applied. This coupling originates from the dependence of the equilibrium position of the oscillator on the charge state of the island and is characterized by the force $F_{st}$ that acts on the
oscillator from the gate electrode when there is an extra electron on it. The coupling can be quantified by a dimensionless constant $\lambda =F_{st}^2/\hbar M \om_0^3$, which is the relative shift of the oscillator energy resulting from a single tunnelling event in respect to zero point motion energy. We assume that oscillator in addition is driven by an external antenna which creates an ac signal with the amplitude $V_d$ and the frequency $\om$. Fig. \ref{scheme} shows the equivalent circuit: The dot represents the island which is connected to four voltage sources via the capacitances. We assume that the displacement of the island $x$ is much smaller then the distance between the island and the gate electrode $d_1$ and both these distances are smaller then the distance between the island and the antenna $d_2$ $(x \ll d1 \ll d2)$. It follows that the gate capacitance is much bigger than the capacitance between the island and the antenna $C_D \ll C_g$. The electrostatic energy of this system in the first order of displacement can be written in the following way,
\begin{equation}
E_{el}=E_{ch}\left(n-\frac{q_0}{e}\right)^2-\frac{C_L V_L^2}{2}-\frac{C_R V_R^2}{2}-\frac{C_g V_g^2}{2}-F(n)x;
\label{el}
\end{equation}
where $E_{ch}=e^2/(C_L+C_R+C_g^0+C_D)$ and $q_0=C_LV_L+C_RV_R+C_g^0V_g+C_DV_D$ are the charging energy and the induced charge on the island by the electrodes respectively ($C_g^0$ is gate capacitance for $x=0$). $n$ represents the charge state of the island, and $F(n)$ is the force acting on the island from all four electrodes, including the driving force. Charge transfer process at a given $n$ is characterized by an energy difference between final and initial state, which is the difference of electrostatic energy given by Eq. (\ref{el}) plus (minus) an energy cost associated with addition (extraction) an electron to (from) the corresponding electrode,
\begin{eqnarray}
\Delta E_{L}^{n \to n+1}&=&E_{ch}\left(n+\frac{1}{2}-\frac{q_0}{e}\right)-eV_L-F_{st}(n)x; \nonumber
\\
\Delta E_{R}^{n+1 \to n}&=&-E_{ch}\left(n+\frac{1}{2}-\frac{q_0}{e}\right)+eV_R-F_{st}(n)x;
\end{eqnarray}
where $F_{st} \equiv F(n+1)-F(n)$.

In what follows, we generalize the treatment of Ref. \onlinecite{Usmani+Blanter07} to the case when the oscillator is ac driven.
At weak coupling, the motion of the oscillator is classical, and our qualitative evaluation of the electron transport in this regime is based on the master equation for the distribution function $P_n (x,v,t)$, where $v$ is the velocity of the oscillator. We assume that the  bias is low enough so that only two charge states, $n$ and $n+1$, are important for the transport. This distribution function obeys the Boltzmann equation \cite{Risken},
\begin{equation}
\label{mastereq}
\left\{\frac{\partial }{\partial t}+v \frac{\partial}{\partial x} + \frac{\partial}{\partial v} \frac{\cal{F}}{M} \right\}P_n = \mbox{St}\ [ P ];\\
\end{equation}
where
\begin{eqnarray}
{\cal{F}} &=& - M \om^2_0 x - M \gamma v+F(n) \; ; \label{totforce}\\
\mbox{St} \ [ P ] &=& (2n-1) \left( \Gamma^{+} (x) P_n
- \Gamma^{-}(x) P_{n+1} \right)+ \nonumber \\
&+&\gamma \left(\frac{\partial}{\partial v}v+\frac{k_B T_{env}}{M} \frac{\partial^2 }{\partial  v^2}\right)P_n. \label{rates2}
\end{eqnarray}
Here $\cal{F}$ is the total force acting on the oscillator which is the sum of the elastic force, the friction force with the damping coefficient $\gamma$ and the electric forces acting from the electrodes. The last ones include the external driving force which we choose to be periodic in time. The "collision integral" $\mbox{St} \ [ P ]$ consists of two terms: the first representing the SET and the second describing interaction of the island with the environment with temperature $T_{env}$. There are four tunnel rates, $\Gamma_{L,R}^{\pm}$, where the subscripts $L$ and $R$ denote tunneling through the left or right junction, and the superscripts $+$ and $-$ correspond to the tunneling to and from the island, respectively. In Eq. (\ref{rates2}) $\Gamma^{\pm} = \Gamma_L^{\pm} + \Gamma_R^{\pm}$. Each rate is a function of the corresponding addition energy $\Delta E_{L,R}^{\pm}$ associated with the addition/removal of an electron to/from the island.

\begin{figure*}[t]
     \centering
		\begin{picture}(2,2)(1,1)
			\put(151,74.5){\includegraphics[width=.11\textwidth]{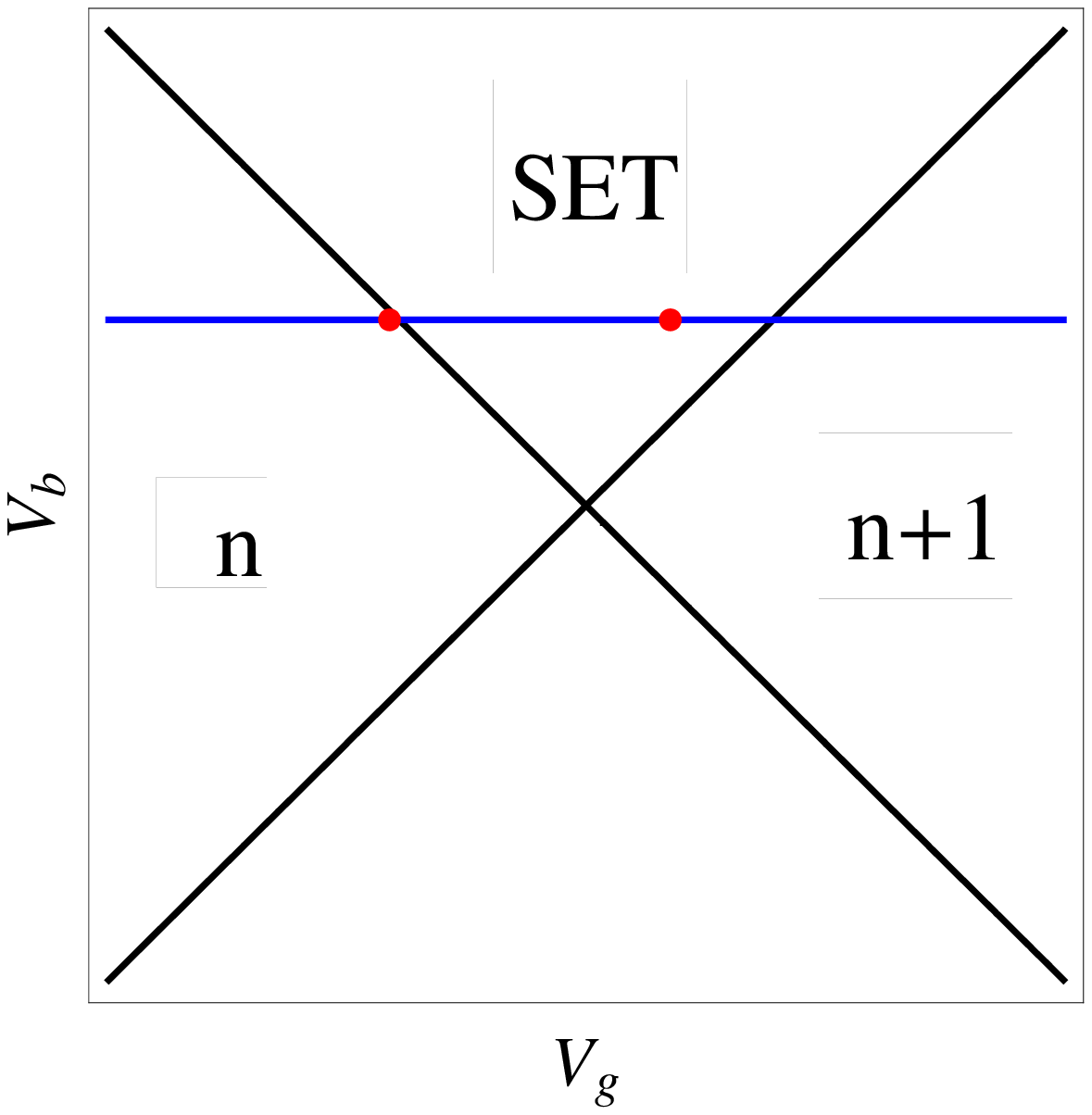}}
		\end{picture}
     \subfigure[]{\label{cur}\includegraphics[width=.41\textwidth]{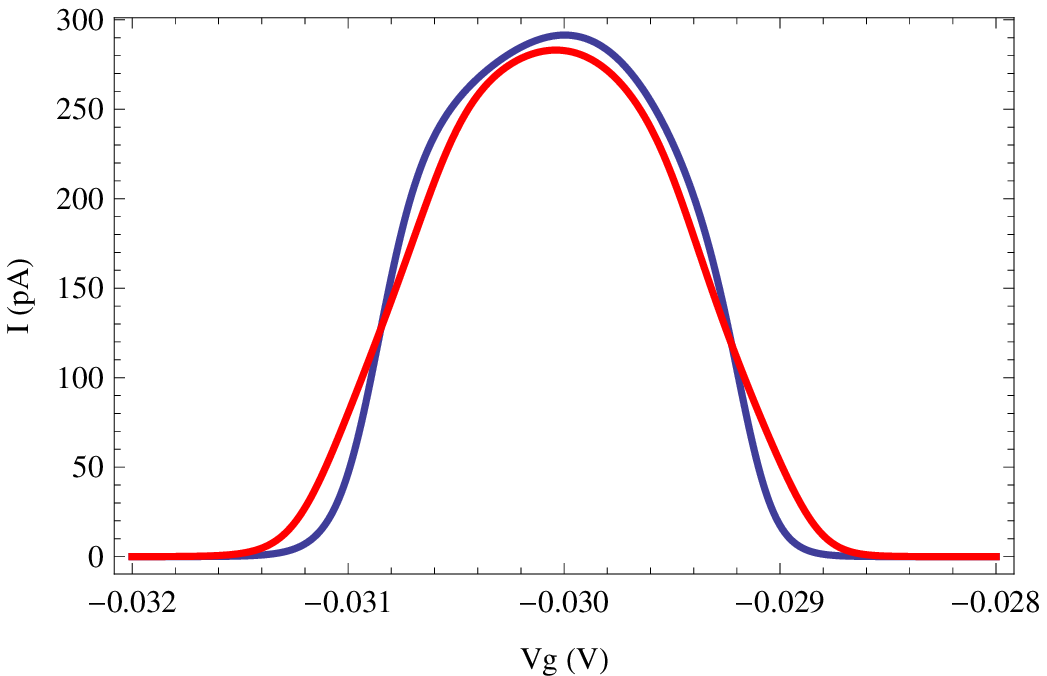}}
     \subfigure[]{\label{res}\includegraphics[width=.35\textwidth, trim = 0mm 0mm 0mm 40mm, clip]{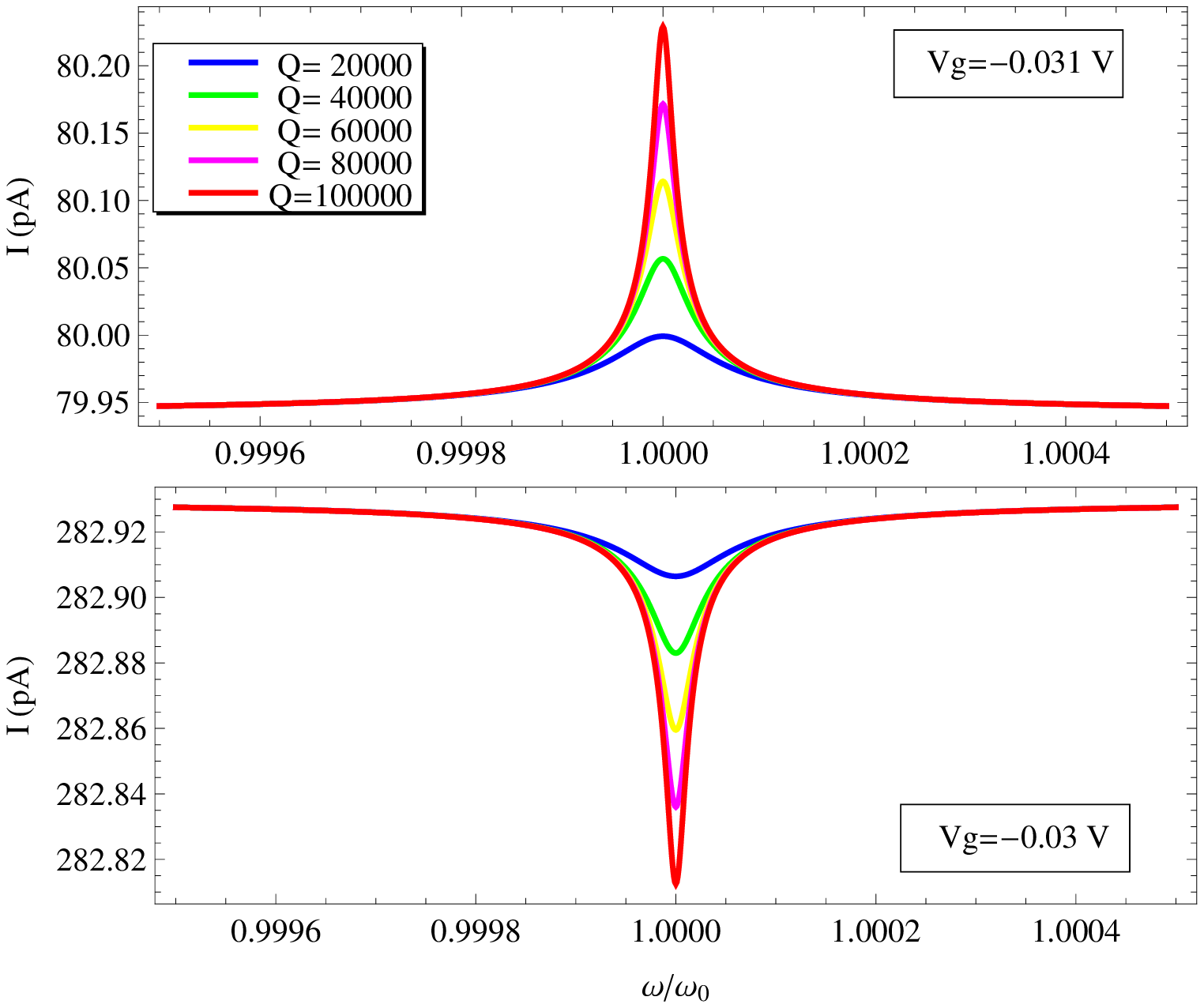}}
     \caption{(a) The I-V characteristics of the system for the fixed value of the bias voltage $V_b=0.35 mV$ and the amplitude of the driving signal $V_d=0.3 V$. The blue (red) curve corresponds to the system out of (at) the resonance. Inset shows the stability diagram, the current in main figure is calculated along the blue line and two red points indicate the parameter values, which were used for calculating the resonances in (b). (b) The current response on the driving frequency for the different values of the quality factor. The upper panel shows the response in the current enhancement region $V_g=-0.031 V$ (resonance) and the lower shows the response in the current suppression region $V_g=-0.03 V$ (unti-resonance)}
     \label{fig2}
\end{figure*}

We can simplify Eq. (\ref{mastereq}) if we restrict ourselves to the fast tunnelling regime $\om \sim \om_0 \ll \Gamma_t$, where  $\Gamma_t = \Gamma^{+} + \Gamma^{-}$ is total tunnelling rate, which means that the oscillator coordinate varies so slowly that $\Gamma(x)$ hardly changes between two successive tunneling events. In this regime we can apply the adiabatic approximation, arriving to the Kramers equation\cite{Usmani+Blanter07},
\begin{equation}
\frac{\partial P}{\partial t} + v \frac{\partial P}{\partial x}+\frac{\partial}{\partial v} \frac{{\cal{F}}_1 P}{M}=A(x)\frac{\partial P}{\partial v}+D(x) \frac{\partial^2 P}{\partial  v^2}+\kappa(x) \frac{\partial (v P)}{\partial v}; \label{kramers}\\
\end{equation}
where ${\cal{F}}_1$ is the total force $\cal{F}$ excluding the friction force. Coefficients $A(x)$, $D(x)$ and $\kappa(x)$ are the following functions of the tunneling rates,
\begin{eqnarray}
D(x)&=&\gamma \frac{k_B T_{env}}{M}+\frac{F^2_{st}}{M^2}\frac{\Gamma^+ \Gamma^-}{\Gamma^3};\label{dif} \\    \kappa(x)&=&\gamma+\frac{F_{st}}{M}\frac{1}{\Gamma}\frac{\partial}{\partial x}\left(\frac{\Gamma^+}{\Gamma}\right); \\
A(x)&=&-\frac{F_{st}}{M}\frac{\Gamma^+}{\Gamma}.
\end{eqnarray}
Here $D(x)$ is the diffusion coefficient which contains the term describing the driving of the oscillator by the stochastic force $F_{st}$. In further calculation we assume that environment temperature sets the lowest energy scale in our system, $k_bT_{env} \ll \min(E_{ch},eV_b)$, so that we drop first term in Eq. (\ref{dif}). Furthermore, $\kappa(x)$ is the drift coefficient and if this is positive the corresponding term can be interpreted as a dissipation. For some tunneling rates and sufficiently high quality factors the drift coefficient can be negative and in this case it gives accumulation of energy. The third term $A(x)$ has a rather trivial effect, it can be included into the oscillator potential energy and just renormalizes the elastic force, not leading to qualitative new effects. Thus we disregard this term in further consideration.

In the single electron transport regime between $n$ and $n+1$ the island is subject to the force which is the sum of two terms: the total electric force $F(n)$ and the stochastic force $F_{st}$ due to the stochastic electron transfers events. The solution of the equation of motion for the oscillator subject to such force can be written in the following way: $x=x_0+\delta x$, where $x_0$ is due to the total electric force and is a non-stochastic function obtained from the solution of the equation of motion. The second term is a stochastic variable which originates from the stochastic force. If we assume that the quality factor $Q$ is sufficiently big ($Q \gg 1$) and we are out of the resonance ($\om_0 \neq \om$), then the last term can be parameterized by the energy $E$ and the phase $\phi$ as follows, $\sqrt{2E/M\om_0}\sin \phi$.

The master equation can be solved for two limiting cases. The first is when $x_0 \gg \delta x$, dependence of the diffusion and feedback coefficients on  $\delta x$ can be disregarded. After averaging Eq. (\ref{kramers}) over the phase we obtain the following,
\begin{equation}
\frac{\partial P}{\partial t}=\frac{\partial}{\partial E}\left[ E \left( \kappa(x_0)+D(x_0)\frac{\partial}{\partial E}\right)P\right]
\end{equation}
We can derive the time averaged distribution function which is
\begin{equation}
P_1(E)=P_0\exp \left[-\frac{\langle \kappa(x_0)\rangle_t}{\langle D(x_0)\rangle_t}E\right]
\end{equation}
Here $\langle \kappa(x_0)\rangle_t$ and $\langle D(x_0)\rangle_t$ are the time averaged drift and diffusion coefficients. If for some values of parameters the drift coefficient becomes negative, the probability increases which means that the oscillator will be driven to higher amplitudes until the energy $E$ hits the value $E_0$, where the system switches to the second regime which corresponds to $x_0 \ll \delta x$. This second regime has been studied in Ref. \onlinecite{Usmani+Blanter07}. The stationary distribution function in this regime is given by
\begin{equation}
P_2(E)=P'_0 \exp\left( - \int_{E_0}^E d E' E' \kappa(E')/ D(E')
\right) \ .
\end{equation}
Here $D(E)=\langle D(x) \cos^2\phi\rangle_\phi$ and $\kappa(E)=\langle \kappa(x) \cos^2\phi\rangle_\phi$,
where the angular brackets denote the average over the oscillation period, $\langle A(\vary)\rangle =\int (d \phi/2 \pi) A(\sqrt{2\epsilon}\sin \phi)$. For the entire energy range, whereas we can not solve the Kramers equation exactly, we can approximate the solution as follows,
\begin{equation}
 P(E) = \left\{ \begin{array}{ll}
         P_1(E) & \mbox{if \ $E < E_0$};\\
        P_2(E) & \mbox{if  \ $E \geq E_0$}.\end{array} \right.
\end{equation}

The average electric current flowing through the system can be calculated by using this distribution function
\begin{equation}
I=\int_0^\infty dE I(E)P(E);
\end{equation}
where
\begin{equation}
I(E)\equiv \frac{e \om}{4 \pi ^2} \int d\phi dt \frac{\Gamma^+_L \Gamma^-_R}{\Gamma_t}.
\end{equation}
Using these equations it is possible to calculate the current flowing through the system if we know the energy dependence of the tunneling rates.

\begin{figure*}[t]
     \centering
		\begin{picture}(2,2)(0,0)
			\put(22,65){\includegraphics[width=.12\textwidth]{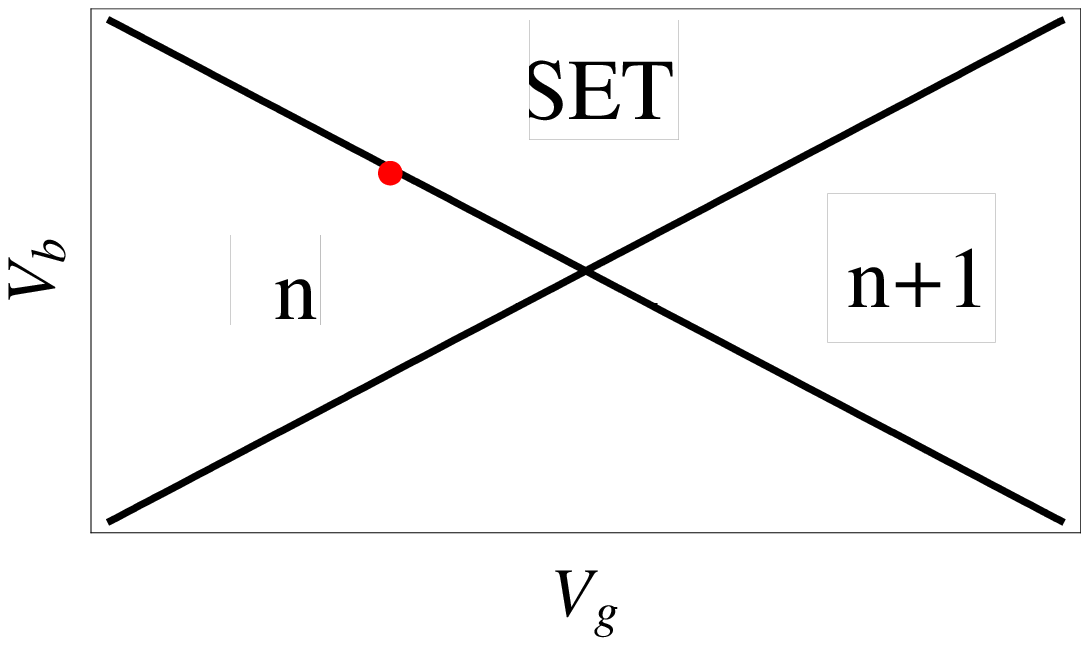}}
		\end{picture}
     \subfigure[]{\includegraphics[width=.3\textwidth]{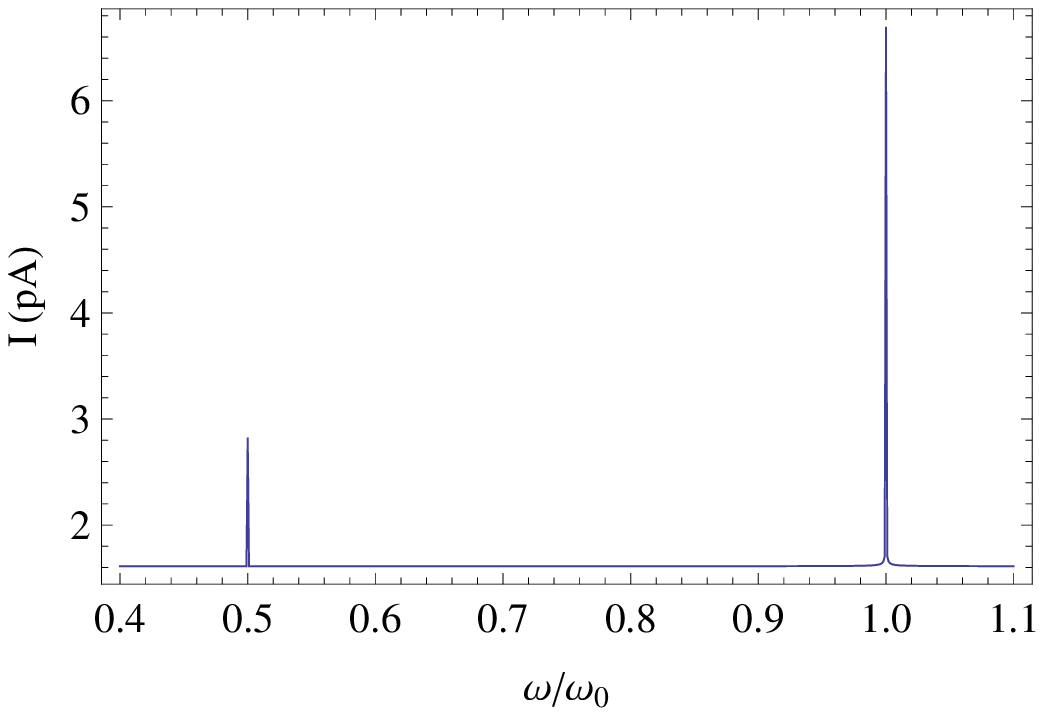}}
		\begin{picture}(2,2)(0,0)
			\put(185,20){\includegraphics[width=.12\textwidth]{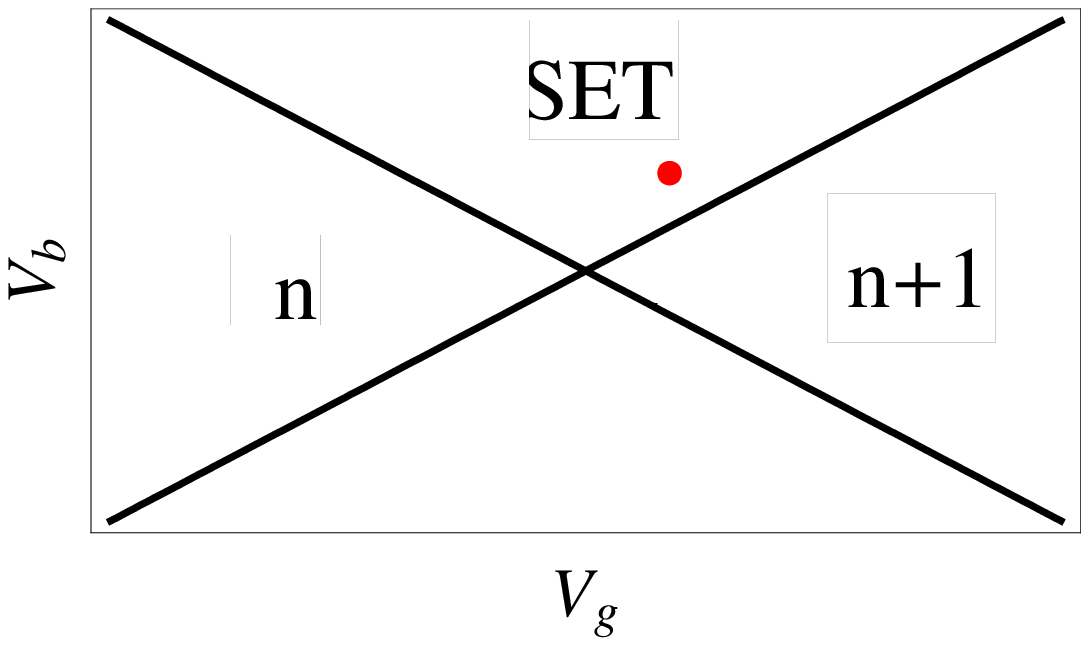}}
		\end{picture}
     \subfigure[]{\includegraphics[width=.3\textwidth]{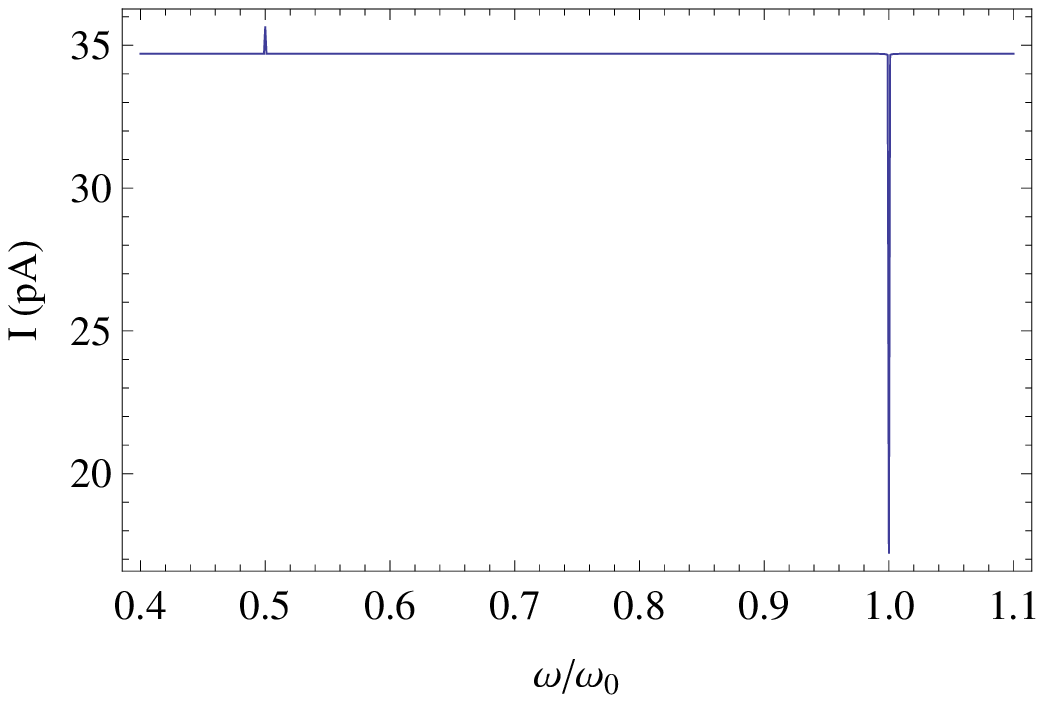}}
		\begin{picture}(2,2)(0,0)
			\put(-134,20){\includegraphics[width=.12\textwidth]{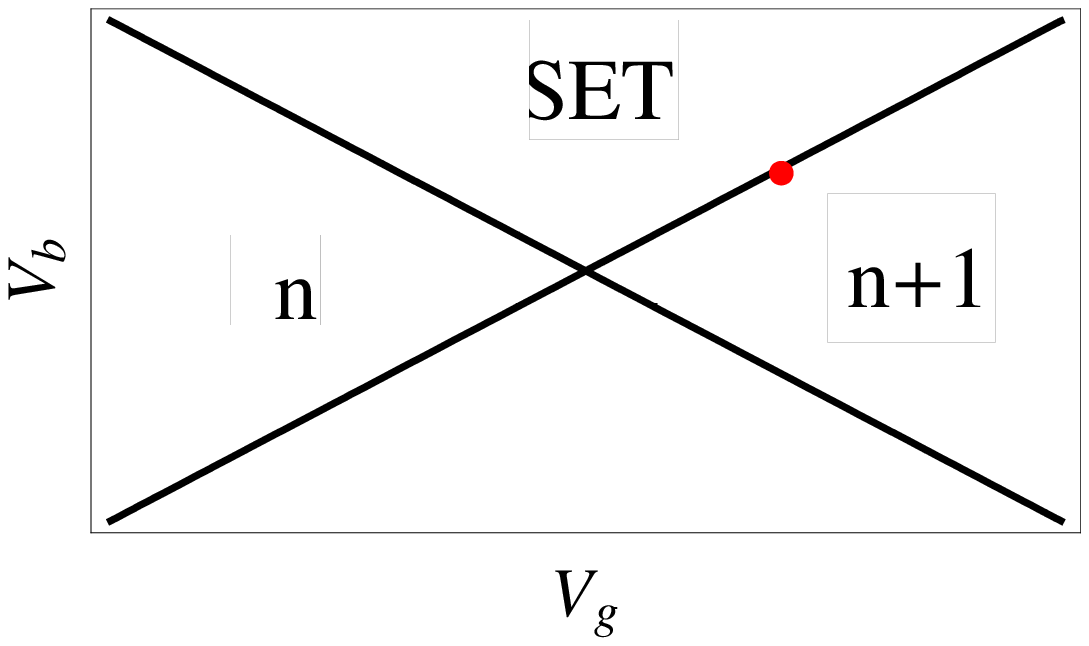}}
		\end{picture}
     \subfigure[]{\includegraphics[width=.3\textwidth]{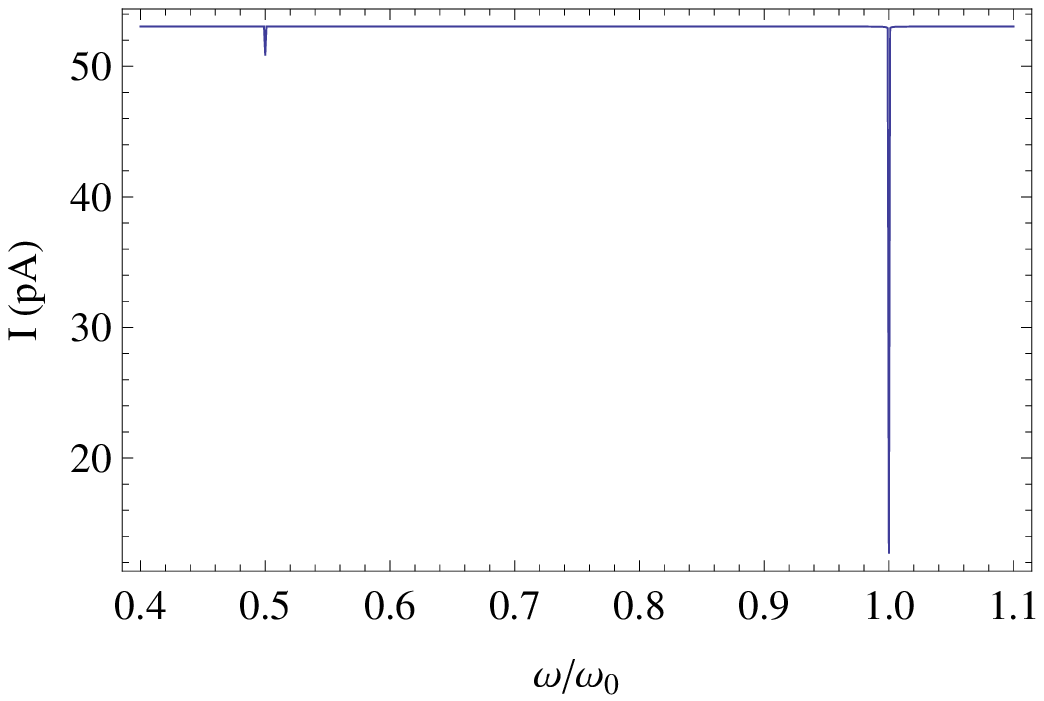}}
     \caption{The dependence of the current on the driving frequency for the fixed values of voltages. Each panel correspond to three different values of gate values: (a) -  $V_g=-0.0315 V$, (b) - $V_g=-0.0296 V$, (c) -$V_g=-0.03 V$. Red points on the insets show parameter values, which were used to calculate the resonances in the main figures.}
     \label{RES}
\end{figure*}

\section{numerical results}

For illustration we have chosen the system considered in Ref. \onlinecite{Zant09}, which is similar to the setup depicted in Fig. \ref{setup-scheme}. The exponential energy dependence of the tunneling barriers is typical for such SET system \cite{Korotkov+Nazarov91}, so we have considered the following rates,
\begin{eqnarray} \label{rates1}
\Gamma_{L,R}^{+} & = & 2b_{L,R} e^{-a_{L,R} \Delta E_{L,R}^{+} }
f_F (\Delta E_{L,R}^{+}) \ ; \nonumber \\
\Gamma_{L,R}^{-} & = & b_{L,R} e^{-a_{L,R} \Delta E_{L,R}^{-}}
f_F (\Delta E_{L,R}^{-}) \ ,
\end{eqnarray}
where $a_{L,R}$ and $b_{L,R}$ are some constants, and $f_F(E)=1/[1+\exp(E/k_bT_{env})]$ is the Fermi function. The factor $2$ accounts for the spin degeneracy of the state $n$. For more concreteness, we take $C_L =C_R =160 aF$ and $C_g =1000C_D=80 aF$. The temperature of the transported electrons we choose to be $T_{env} = 200 mK$ and the value of the oscillator frequency $\om_0$ we take to be 1 GHz.

\begin{figure}[b]
\centering
\includegraphics[scale=0.75]{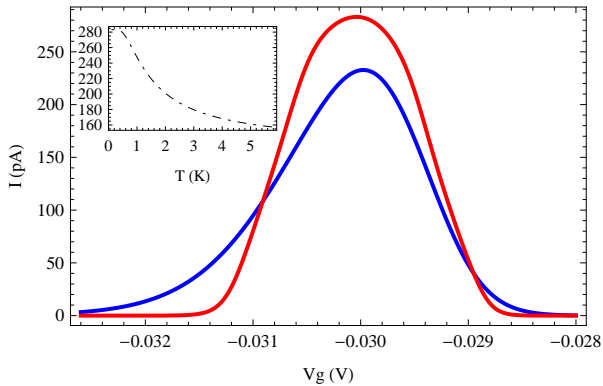}
\caption{The current-voltage characteristics of the system for two different values of temperature of tunneling electrons: red correspondes to $T=0.2 K$ and blue-$T=1 K$. Inset showes dependence of current on the temperature of the electrons for fixed values of the electrod voltages: $V_g=-0.0315 V$, $V_b=0.35 mV$ and $V_d=0.3 V$ }\label{temps}
\end{figure}

Fig. \ref {cur} represents the current-voltage characteristics for the above described system in the single electron transport regime. For the fixed value of bias voltage $V_b=0.35 mV$ we have varied the gate voltage and have calculated the current. The blue curve on Fig. \ref{cur} corresponds to the immovable island. When the island is in resonance with the driving signal, which we take with the amplitude $V_d=0.3 V$, the current is modified due to the oscillations (red curve). For the different gate voltages the current at the resonance is either suppressed or enhanced. The linear response of the current on the driving frequency is shown on the figure \ref{res} for the different values of the quality factor. Depending on which region we take, the current enhancement or the current suppression, we get the linear resonance or anti-resonance peaks.

The electric force $F(n)$ in Eq. (\ref{el}), acting on the island from the electrodes can be decomposed as the sum of three terms:
\begin{equation}
F(n)=f_1(V,N)cos^2(\om t)+f_2(V,N)cos(\om t)+f_3(V,N)
\label{force}
\end{equation}
where $f_1(V,N)$, $f_2(V,N)$ and $f_3(V,N)$ depend on the charge state of the island and the voltages applied on the electrodes (bias, gate and driving voltages). Because of the first term in Eq. (\ref{force}) the amplitude of the oscillations increases at the half of the driving frequency $\om=\om_0/2$. This gives an additional response peak of the current at this value of the frequency.

Fig. \ref{RES} shows the current as the function of the driving frequency for the fixed values of voltages ($V_b=0.35 mV$, $V_d=7 V$). Three different panels correspond to the different values of the gate voltage. Thus, by changing the gate voltage it is possible to get either two resonance peaks (c), two anti resonance peaks (b) or one resonance and one anti-resonance peaks (a) at the resonance frequency and at it's half.

Increasing the temperature of transferred electrons causes broadening of coulomb picks and coulomb diamonds. Fig. \ref{temps} shows current dependence on gate voltage for different values of electron temperature. For higher values of the temperature the current pick decreases in value and broadens (blue curve). Inset represents temperature dependence of the current for fixed values of applied voltages, which shows that for high temperatures the height of Coulomb peak decreases exponentially.

\section{conclusion}
In this Article, we studied the behavior of the "self-detecting" NEMS in the regime were it can modeled as a single electron transistor weakly coupled to a harmonic mechanical oscillator. In particular, we studied the dependence of electric current on the external driving frequency. We demonstrated that current is modified when the system is at the resonance. We also found that since current can be enhanced by mechanical motion as well as suppressed, depending on the applied voltages, the driving of the resonator can result in the resonance peak (associated with the current enhancement) as well as in an antiresonance dip (associated with the current suppression). Additionally, the driving force is proportional contains doubly-oscillating terms, which results in the appearance of an additional resonance peak/dip in the spectrum at the half of the natural frequency $\omega_0$. Depending on the applied voltages, both resonances can be peaks, dips, or one can have a peak and a dip. The latter situation is realized provided the current is enhanced by mechanical motion at low amplitudes and suppressed at high amplitudes, or vice versa. The regions of strong backaction discussed in Ref. \onlinecite{Usmani+Blanter07} for the system studied above appear to be for high, where the system is not in the single electron tunnelling regime anymore, and our simple analysis can not be applied to this situation.

This Article treats the simplest situation --- when the oscillator is linear, and the coupling is weak. Experiments \cite{Zant09} clearly indicate that in certain voltage and power range the resonator can be driven into non-linear and/or strong coupling regime\cite{strong}. With the increase of the driving power the oscillator enters non-linear regime, and bistabilities and switching effects between states with different amplitude have been experimentally observed. Current response to the external frequency in a non-linear system will have a more complicated form that discussed in this Article, in particular, resonances at the frequencies multiple of $\omega_0/2$ are expected. These effects require a more extensive analysis which will be performed elsewhere.

\section{acknowledgments}
We thank  H. S. J. van der Zant, H. Meerwaldt, M. Poot and C.J.O. Verzijl for useful discussions. This work was supported by the Netherlands Foundation for Fundamental Research on Matter (FOM).

\end{document}